\documentclass[a4paper,11pt]{article}
\usepackage{jheppub} % for details on the use of the package, please
\usepackage[T1]{fontenc} % if needed
\usepackage{CJK}
\usepackage{setspace}
\usepackage{float}
\def\be{\begin{equation}}
\def\ee{\end{equation}}
\def\bea{\begin{array}}
\def\eea{\end{array}}
\def\beqa{\begin{eqnarray}}
\def\eeqa{\end{eqnarray}}
\def\beqas{\begin{eqnarray*}}
\def\eeqas{\end{eqnarray*}}

\def\bp{\begin{picture}}
\def\ep{\end{picture}}
\def\bc{\begin{center}}
\def\ec{\end{center}}
\def\bfig{\begin{figure}}
\def\efig{\end{figure}}

\def\bit{\begin{itemize}}
\def\eit{\end{itemize}}
\def\nn{\nonumber}
\def\f{\frac}

\def\[{\left[}
\def\]{\right]}
\def\({\left(}
\def\){\right)}

\def\..{\left.}
\def\.{\right.}
\def\tl{\tilde}
\def\ra{\rightarrow}
\def\la{\leftarrow}

\def\tm{\times}

\def\da{\dagger}

\def\la{\lambda}

\def\al{\alpha}

\def\ep{\epsilon}

\def\pa{\partial}

\title{Radiative natural SUSY spectrum from deflected AMSB scenario with messenger-matter interactions}
\author[a,b]{Fei Wang,}
\author[b,c]{Jin Min Yang,}
\author[b]{Yang Zhang}

\affiliation[a]{School of Physics, Zhengzhou University, Zhengzhou 450000, P. R. China}
\affiliation[b]{State Key Laboratory of Theoretical Physics, Institute of Theoretical
                Physics, Chinese Academy of Sciences, Beijing 100080, P. R. China}
\affiliation[c]{Department of Physics, Tohoku University, Sendai 980-8578, Japan}

\emailAdd{feiwang@zzu.edu.cn}
\emailAdd{jmyang@itp.ac.cn}
\emailAdd{zhangyang@itp.ac.cn}

\abstract{A radiative natural SUSY spectrum are proposed in the deflected anomaly mediation scenario with general messenger-matter
interactions. Due to the contributions from the new interactions, positive slepton
masses as well as a large $|A_t|$ term can naturally be obtained with either sign of
deflection parameter and few messenger species (thus avoid the possible Landau pole
problem). In this scenario, in contrast to the ordinary (radiative) natural SUSY
scenario with under-abundance of dark matter (DM), the DM can be the mixed bino-higgsino
and have the right relic density. The 125 GeV Higgs mass can also be easily obtained
in our scenario. The majority of low EW fine tuning points can be covered by the XENON-1T direct detection experiments.
}

\begin{document}
\maketitle

\newpage
\section{Introduction}
The standard model (SM) of particle physics has been confirmed by various experiments.
Especially, a 125 GeV SM-like Higgs boson was discovered by both the ATLAS \cite{ATLAS:higgs}
and CMS collaborations \cite{CMS:higgs} of the Large Hadron Collider (LHC). On the other hand,
the SM, as a successful effective theory, has many theoretical or aesthetical problems which
necessitate various extensions. Low energy supersymmetry (SUSY) is a highly motivated paradigm
for physics beyond the SM. In fact, an interesting observation is that the Higgs mass lies
miraculously in the narrow $115-135$ GeV window predicted by the minimal SUSY model (MSSM).
In addition, the top quark mass also lies exactly at what is needed to properly drive the
radiative electroweak symmetry breaking (EWSB). Besides, the gauge hierarchy problem,
the successful gauge coupling unification requirement as well as the dark matter (DM) puzzle
can all be solved by SUSY.

The low energy SUSY paradigm is appealing, but so far there is no sign of SUSY particles after
extensive searches at the LHC. In fact, no significant deviations from the SM have been observed
in electroweak precision measurements as well as in flavor physics. The LHC data has already
set stringent constraints \cite{CMSSM1,CMSSM2} on certain CMSSM models:
$m_{\tilde g} \gtrsim 1.8$ TeV for $m_{\tl{q}} \sim m_{\tl{g}}$, and $m_{\tl{g}}\gtrsim 1.3$ TeV
for $m_{\tl{q}} \gg m_{\tl{g}}$.
Besides, the rather large value of the Higgs mass at 125 GeV requires TeV-scale highly mixed
top squarks, which seems to contradict to the expectation from naturalness. In order to generate
a soft SUSY spectrum that can be consistent with the LHC discoveries, a proper SUSY breaking
mechanism is needed.

One of the most elegant SUSY breaking mechanisms is the anomaly mediation \cite{AMSB} SUSY breaking
scenario.  The ordinary AMSB has many advantages and is very predictive. However, it has the
tachyonic slepton problem \cite{tachyonslepton} and needs an extension.  An elegant extension
to tackle the tachyonic slepton problem is the deflected  AMSB scenario \cite{deflect},
in which the messengers are introduced to deflect the Renormalization Group Equation (RGE)
trajectory. The tachyonic slepton problem can be solved with such a deflection. On the other hand,
many messenger species are needed to push slepton masses positive for a negative deflection
parameter. A large number of messenger species could cause the Landau pole below the Planck scale.
Besides, a large fine-tuning is needed to generate the 125 GeV Higgs mass in the ordinary
deflected AMSB scenario.

In our previous work \cite{fei:PLB}, we proposed to introduce general messenger-matter interactions
in the deflected AMSB scenario. The slepton sector can receive additional contributions from both
the messenger-matter interactions and the ordinary deflected anomaly mediation to avoid tachyonic
slepton masses. At the same time, additional contributions to trilinear coupling $A_t$ term which
typically increase $|\tilde{A}_{t}|(\equiv A_t-\mu\cot\beta|)$ could be helpful to give the 125 GeV
Higgs and reduce the fine-tuning involved. Besides, even with one messenger we can generate
positive slepton masses regardless the sign of deflection parameters \cite{okada}.
So the Landau pole problem can be evaded in our new scenario.

Note that with a large $A_t$ term and the TeV-scale stops as well as a small
$\mu\sim 100-300 {\rm GeV}$, the radiative natural SUSY scenario \cite{naturalsusy,rnaturalsusy}
can naturally be realized in the deflected AMSB with general messenger-matter interactions.
The electroweak (EW) fine-tuning \cite{EWFT} is small (typically $\Delta_{EW}<50$), especially
when $A_t$ is large which will
decrease the fine-tuning involved. On the other hand, the DM in ordinary natural SUSY will
always be higgsino-like and results in under-abundance. Although two-component dark matter
(axion and higgsino) can work well \cite{Baer:axion}, it is preferable to change the nature
of DM. We know that the gaugino mass relation in the ordinary AMSB is different from
the relation in gauge mediation and gravity mediation. It will result in wino-higgsino
DM and thus the under-abundance problem persists \cite{Baer:nugaugino}. With the deflection
of AMSB trajectory, the DM can be the mixed bino-higgsino and could give the right relic density.
In this work we focus on such a realization of the radiative natural SUSY in
 the deflected AMSB with general messenger-matter interactions.

This paper is organized as follows. We briefly review the deflected AMSB scenario with
general messenger-matter interactions in Sec. 2. In Sec. 3 we introduce new messenger-matter
interactions to the deflected AMSB and study the soft parameters which can generate
the radiative natural SUSY.  Numerical results are presented in Sec. 4.
Sec. 5 contains our conclusions.

\section{A review on deflected AMSB with matter-messenger interactions}
We briefly review the general results of the deflected AMSB scenario with general matter-messenger
interactions. The relevant details can be found in our previous study \cite{fei:PLB}.
General messenger-matter interactions in GMSB can be seen in various papers \cite{chacko,shih,gmsb-mm}. 

The superpotential in the deflected AMSB scenario includes general messenger-matter interactions:
 \beqa
 W=\la_{\phi ij}X Q_iQ_j+ y_{ijk}Q_i Q_j Q_k+W(X)~,
 \eeqa
where the indices $'i,j'$ run over all MSSM and messenger fields. Subscripts $'U,D'$ will denote
the cases up and below the messenger threshold, respectively. $W(X)$ denotes the superpotential for pseduo-moduli field $X$ which defines the messenger threshold.

After integrating out the messenger fields, we have the general form for the MSSM fields only:
\beqa
{\cal L}=\int d^4\theta  Q_a^\da Z_D^{ab}(\f{\mu}{\sqrt{\phi^\da\phi}},\sqrt{\f{X^\da X}{\phi^\da \phi}}) Q_b+\int d^2\theta y_{abc}Q^aQ^bQ^c~,
\eeqa
which can give additional contributions to soft SUSY breaking parameters. Here $'\phi'$ denotes
the compensator field with Weyl weight 1 and $'Z_D'$ the wavefunction renormalization factor below the messenger threshold.

The leading-order contributions to the trilinear terms and scalar terms are
\beqa
\f{A_{abc}}{y_{abc}}&=&\sum\limits_{i=a,b,c}\left(-\f{1}{2}F_\phi\f{\pa}{\pa\ln \mu} +\f{d F_{\phi}}{2}\f{\pa }{\pa \ln |\tl{X}|}\right)\ln Z^{ii}_D({\mu},|\tl{X}|)~, \\
m^2_{ab}&=&\left(-\f{|F_\phi|^2}{4}\f{\pa^2 }{\pa (\ln\mu)^2}-\f{|F_{\tl{X}}|^2}{4}\f{\pa^2}{\pa |\tl{X}|^2}+\f{|F_\phi||F_{\tl{X}}|}{2}\f{\pa^2}{\pa\ln\mu\pa|\tl{X}|}\right)\ln Z_D^{ab}(\mu,|\tl{X}|)~,\nn\\
&=&\left[-\f{|F_\phi|^2}{4}\f{\pa^2}{\pa (\ln\mu)^2}-\f{d^2|F_{\phi}|^2}{4}\f{\pa^2}{\pa \ln|\tl{X}|^2}+\f{d|F_\phi|^2}{2}\f{\pa^2}{\pa\ln\mu\pa\ln|\tl{X}|}\right]\ln Z_D^{ab}(\mu,|\tl{X}|)~,
\eeqa
where the last term is the unique feature of this deflected AMSB scenario which involves
the interference between the pure anomaly and gauge mediation type contributions.

Following the conventions in \cite{shih}, the derivative of the wavefunction with respect to $t=\ln\mu$ are given as
\beqa
\label{anomalous}
\f{d Z_{ij}}{dt}&\equiv& G_{ij}[Z(\ln\mu);\la(\ln\mu);g(\ln\mu)]~,\nn\\
    &=&-\f{1}{8\pi^2}\left(\f{1}{2}d_i^{kl}\la_{ikl}^* Z_{km}^{-1~*}Z_{ln}^{-1~*}\la_{jmn}-2c_r^i Z_{ij}g_r^2\right)~,
\eeqa
we can obtain the expression for the first derivative of wavefunction with respect to $'X'$ \cite{chacko} at the messenger scale $\mu=|X|$
\beqa
\f{\pa Z_D^{ab}(\ln\mu,|X|)}{\pa X}&=&\f{1}{2X}\Delta G^{ab}~,
\eeqa
with $\Delta(\cdots )$ denoting the discontinuity of its followed expression,
and $d_i^{kl}$ being the standard multiplicity factor in the one-loop anomalous dimensions.

The interference terms between the anomaly mediation and gauge mediation are
\beqa
&& \f{\pa^2}{\pa\ln\mu\pa \ln |\tl{X}|} Z_D^a(\mu,|\tl{X}|)=\f{\pa}{\pa \ln|\tl{X}|}G_a[Z_D(\ln\mu,\tl{X});\la(\ln\mu,\tl{X});g(\ln\mu,\tl{X})]~,\nn \\
&&=\(\Delta(\beta_{\la})\f{\pa}{\pa \la}+\Delta(\beta_g)\f{\pa}{\pa g}+\f{\pa Z_D^a}{\pa\ln\tl{X}}\f{\pa}{\pa Z_D^a}\)G_a[Z_D^a(\ln\mu);\la(\ln\mu);g(\ln\mu)]~.
\eeqa
So we arrive at the final results for the trilinear and scalar soft masses with a general
messenger sector at the messenger scale \cite{fei:PLB}:
\beqa
&& A_{a}=-\f{1}{2}G^D_{aa} F_\phi-\f{1}{32\pi^2} d_a^{ij}\Delta(|\la_{aij}|^2) d F_\phi~,\\
&& m^2=m^2_{\rm AMSB}+m^2_{\rm gauge}+m^2_{\rm inter}~,
\label{all}
\eeqa
with
\beqa
m^2_{\rm AMSB}&=&\left[-\f{|F_\phi|^2}{4}\(\f{\pa \gamma^a }{\pa g_i}\beta(g_i)+\f{\pa \gamma^a}{\pa y_i}\beta(y_i)\)\right]~,\\
(m^2_{ab})_{\rm inter}&=&\f{d F_\phi^2}{2}\left\{\f{}{}
-\f{1}{8\pi^2}\[d_a^{kl}\la_{akl}^*\la_{bmn}(\f{\Delta G^D_{km}}{2}+\f{\Delta G^D_{ln}}{2})+2 c_r^i g_r^2\f{\Delta G^D_{ab}}{2}\]\right.\nn\\
%%%%&-&\f{1}{8\pi^2} d_i^{kl}\la_{ikl}^* {\Delta (\beta_\la)}
&+&\left.\f{1}{8\pi^2} 4 c_r^k \f{1}{16\pi^2} g_k^4  {\Delta} ({b_k})-G_D\f{\Delta G_D}{2}\right\}~,
\eeqa
and the gauge mediation type contributions similar to \cite{shih}:
\beqa
\label{GMSB}
(m^2_{ab})_{\rm gauge}&=&\f{d^2 F_\phi^2}{4}\f{1}{256\pi^4}\left[\f{1}{2}d_a^{ik}d_{i}^{lm}\(\Delta(\la_{aik}^*\la_{bjk})(\la_{ilm}\la_{jlm}^*)^U\)
-(\la_{aik}^*\la_{bjk})^D\Delta(\la_{ilm}\la_{jlm}^*)\nn\right.\\
&+&\left. \f{1}{4}d_{a}^{ij}d_b^{kl}\Delta(\la_{aij}^*\la_{cij})\Delta(\la_{ckl}^*\la_{bkl})-d_a^{ij}C_r^{aij}g_r^2\Delta(\la_{aij}^*\la_{bij})\right].
\eeqa

\section{Deflected AMSB with new messenger-matter interactions}
The characteristic feature of the radiative natural SUSY with respect to the ordinary natural SUSY
is the large $|A_t|$ term. In order to obtain the relatively large trilinear terms,
we include new messenger-matter interactions in the deflected AMSB scenario.
The messengers are introduced in pairs of $({\bf 5},\bar{\bf 5})$ representations of SU(5).
So the messengers obviously have the following decomposition in terms of the SM
$SU(3)_c\tm SU(2)_L\tm U(1)_Y$ quantum number:
 \beqa
 {\bf 5}_a&=& N_a(1,2)_{1/2}\oplus M_a(3,1)_{-1/3}~,\nn\\
 \bar{\bf 5}_a&=&\overline{N}_a(1,\bar{2})_{-1/2}\oplus\overline{M}_a(\bar{3},1)_{1/3}~,
 \eeqa
with $'a'$ denoting the $N_F$ messenger species.

 We introduce the following superpotential that involves the messenger-MSSM interaction
 \beqa
 W^U&\supset& X \overline{N}_a N_a + X\overline{M}_a M_a+W(X)\nn\\
  &+&\sum\limits_{i}\[\la^D_{ai} Q_L^i (D_R^c)_i \overline{N}_a +\la^L_{ai} L_i (E_L^c)_i\overline{N}_a
 + \la^U_{ai} Q_L^i (U_R^c)_i {N}_a \]~,\nn
 \eeqa
with the typical form of superpotential $W(X)$ for pseduo-moduli field $X$ to determine
the deflection parameter $d$ in combination with $F_\phi$. Here the superscript $'i'$
denotes the family indices.

From the general expressions of soft parameters in Sec. 2, we can obtain the soft SUSY breaking
parameters for sfermions and trilinear couplings at the messenger scale. We keep the leading
terms involving only $y_t,g_3,\la_{L,i},\la_{U,i},\la_{D,i}$. Subleading terms like
$g_{1,2}^4,y_{b,\tau}^2\la_{L,U,D;ia}^2$ are not kept in the following expressions.
For simplicity, we set family and messenger species universal couplings
$\la_{L,ai}=\la_L; \la_{U,ai}=\la_U;\la_{D,ai}=\la_D$ for messenger-matter interactions.
Besides, we only give explicitly the soft terms for the third generation squarks.
The first two generation squarks can be obtained by removing the $y_t^2$ terms in the
relevant expressions. The soft SUSY mass terms for the three generations of sleptons
have the same form. The values of $\mu$ and $B\mu$ are model-dependent and we leave them
as free parameters because we do not give an explicit mechanism in our scenario.
They are determined by successful EWSB conditions.

The gaugino masses are given by
 \beqa
 M_i&=&-\f{\al_i}{4\pi}(b_i+N_F d)~,
 \eeqa
 with the beta function $(b_1,b_2,b_3)=(33/5,~1,-3)$ and the standard normalization
for $g_1$ coupling $g_1^2=5g_Y^2/3$.

The trilinear couplings are calculated to be
\beqa
\label{AT}
A_t&=&\f{F_\phi}{16\pi^2}\[6y_t^2-(3\la_U^2+\la_D^2)d-\f{16}{3}g_3^2\]~,\nn\\
%%%%16 g_3^2/3=7.372
A_b&=&\f{F_\phi}{16\pi^2}\[y_t^2-(\la_U^2+3\la_D^2)d-\f{16}{3}g_3^2\]~,\nn\\
A_\tau&=&\f{F_\phi}{16\pi^2}\(-3\la_L^2d\)~.
\eeqa
The soft parameters are
\beqa
\f{m^2}{F_\phi^2}&=&\f{d}{2}\delta^m+\f{d^2}{4}(\delta^G+\delta^3)+\f{1}{4}\delta^A,
\eeqa
with the relevant tedious expressions given in the appendix.

We have the following discussions:
\bit
\item[(i)]  In our scenario, the notorious tachyonic slepton problem which appears in the ordinary
AMSB can be naturally solved. Besides, the slepton masses receive (dominant) positive contributions
from matter-messenger interactions regardless of the sign of the deflection parameter $d$.

\item[(ii)]  In our scenario, even one messenger specie can work well to give positive
slepton masses regardless of the sign of deflection parameter $d$. So the possible Landau
pole problem below the Planck scale will naturally be evaded in our scenario.

\item[(iii)] The $A_t$ value can be either positive or negative, depending on the sign of $d$.
Large $\la_{U,D}$ can lead to a large value of $|A_t|$ which can naturally give a large Higgs
mass with a less fine-tuning.

\item[(iv)]  There is some parameter space for light soft stop masses. So the radiative natural
SUSY spectrum can be realized in our scenario. We will discuss such a realization in next section.

\eit

\section{Radiative natural SUSY spectrum and numerical analysis}
The 125 GeV Higgs has already set some constraints on the low energy SUSY spectrum.
Obviously from the formula
\beqa
m_{h}^{2}&\simeq& m_{Z}^{2}\cos^{2}2\beta+\frac{3m_{t}^{4}}{4\pi^{2}v^{2}}
\left[\log\frac{M_{\mathrm{SUSY}}^{2}}{m_{t}^{2}}+\frac{\tilde{A}_{t}^{2}}{M_{\mathrm{SUSY}}^{2}}\left(1-\frac{\tilde{A}_{t}^{2}}{12M_{\mathrm{SUSY}}^{2}}\right)\right],
\eeqa
with
\beqa
\tl{A}_t=A_t-\mu\cot\beta,~~~~M_{\mathrm{SUSY}}^2=m_{\tilde{t}_{1}}m_{\tilde{t}_{2}}~,\nn
\eeqa
we need either $M_{\mathrm{SUSY}}/m_{t}\gg1$ or $M_{\mathrm{SUSY}}/m_{t}>1$ with
$\tilde{A}_{t}/M_{\mathrm{SUSY}}>1$.  The stop masses must be larger than 10 TeV in case of
no stop mixing, and hence a large fine tuning is needed.  Obviously, a large $\tl{A}_t$ is
preferable for low energy SUSY.

The models of natural SUSY \cite{naturalsusy} try to retain the naturalness of weak scale SUSY
by proposing a spectrum of light higgsinos $|\mu|\sim 100-300$ GeV  and
light $\tl{t}_{1,2},\tl{b}_1$ along with very heavy masses of other squarks
and TeV-scale gluinos. The gluino mass can affect the stop masses via RGE evolution.
So, a low EW fine-tuning requires that the gluino mass can not be too heavy.
 On the other hand, it is also bounded from below to be $m_{\tl{g}}  \gtrsim 1.3$ TeV by
the LHC searches within the context of SUSY models like mSUGRA/CMSSM.
The first two generation sfermions can be allowed to lie in the  5-20 TeV range without
introducing unnaturalness.
Heavier first two generation squarks can ameliorates the SUSY flavour, CP, gravitino and
proton-decay problems due to decoupling.
Such models have a low electroweak fine-tuning and satisfy the LHC constraints.

However, the relatively heavy (125 GeV) Higgs
mass has some tension with the ordinary natural SUSY scenario
and indicates that natural SUSY may take the form of radiative natural SUSY \cite{rnaturalsusy}
which requires a large $A_t$ term. In fact, a large $|A_t|$ value can suppress the top squark
contributions to $\Sigma_u^u$ and at the same time lift up the Higgs mass. Such a large $|A_t|$
can easily be obtained in our scenario. We can see from Eq.(\ref{AT}) that a large $|A_t|$
will appear in case of a large $\la$ and either sign of deflection parameter $d$.

In the ordinary radiative natural SUSY scenario with universal gaugino mass at the GUT scale,
the lightest sparticle (LSP) is always the higgsino which can not fully account for the DM
relic abundance. The gaugino relation at the EW scale can naturally be evaded in the deflected
AMSB scenario and thus the DM can be the mixed bino-higgsino or wino-higgsino
(or pure bino, pure wino). We know that in the ordinary AMSB, the gaugino mass ratio at
the EW scale is
\beqas
M_1:M_2:M_3\approx3.29:1:-9.6.
\eeqas
 This can lead to the mixed higgsino-wino dark matter for gluino at about 2 TeV.
As noted in \cite{Baer:nugaugino}, the under-abundance problem of DM persists.
In general, in order to get the mixed higgsino-electroweakino DM, we need the gaugino mass
ratio to satisfy
\beqas
M_3: \min(M_1,M_2)\gtrsim 5,
\eeqas
with gluino mass heavier than 1.5 TeV.
The mixed bino-higgsino DM can give the full DM abundance.
This prefer a negative deflection parameter with $N_F d\lesssim-3$.

In our scenario, the soft terms are characterized by the following free parameters
\beqa
  N_F, d, \mu,M_{mess}, F_\phi, \tan\beta, \la_U,\la_D,\la_L.
\eeqa
We scan the parameter space with the following messenger scale($M_{mess}$) inputs:
\bit
\item  The $\mu$ parameter is chosen to lie between  $|\mu|\sim 100-300 $ GeV to keep EW naturalness.
\item  The scale of $F_\phi$ determines the whole SUSY spectrum.
 The gaugino masses, the EWSB condition as well the Higgs mass constrain the value of $F_\phi$
to be in the range $10 {\rm TeV}<F_\phi<500 {\rm TeV}$.
\item The messenger scale $M_{mess}$ can be chosen to lie between the GUT scale and the typical sparticle scale:
 $10 {\rm TeV}<M_{mess}<10^{16} {\rm GeV}$.
\item  The value of $\tan\beta$ is chosen to be $40\geq\tan\beta\geq 2$. The messenger species $N_F$ should
lie in the range $1\leq N_F\leq 3$ to avoid the possible Landau pole while the deflection parameter $d$
is chosen to satisfy $N_F\cdot d\lesssim-3$ to fully account for the DM relic density.

\item  For simplicity, we set $\la_U=\la_D=\la$.  We set the range of the messenger-matter interactions:
$0.5\lesssim \la,\la_L \lesssim 3$ to justify our keeping of the leading contributions in previous
calculations and at the same time avoid the possible Landau pole.
\eit
In our scan  we take into account the following collider and dark matter constraints:
\bit
\item[(1)] Successful radiative EWSB condition.
\item[(2)] The stop and sbottom masses can be relatively heavy in the radiative natural SUSY scenario
in contrast to the upper bound of 1.5 TeV in ordinary natural SUSY (with less than 10\% EW fine tuning ).
We require that the stop masses to satisfy $m_{\tl{t}_{1,2}}\lesssim 4 $ TeV which corresponds to
to an upper bound for the EW fine-tuning $\Delta_{EW}\lesssim 50$. A large $|A_t|$ will always decrease
the fine-tuning involved. Due to the gluino loop contribution to the stop masses, the gluino is bounded
to be below 12 TeV.
\item[(3)] The lower bounds on neutralino and chargino masses from LEP, including the invisible decay
of $Z$-boson.
The most stringent constraints of LEP come from the chargino mass and the invisible $Z$-boson decay.
We require $m_{\tl{\chi}^\pm}> 103.5 {\rm GeV}$ and the invisible decay
width $\Gamma(Z\ra \tl{\chi}_0\tl{\chi}_0)<1.71~{\rm MeV}$,
which is consistent with the $2\sigma$ precision EW measurement $\Gamma^{non-SM}_{inv}< 2.0~{\rm MeV}$.
\item[(4)] The combined mass range for the Higgs boson: $123 {\rm GeV}<M_h <127 {\rm GeV}$
from ATLAS and CMS data \cite{ATLAS:higgs,CMS:higgs}.
\item[(5)] The relic density of the neutralino dark matter satisfies the Planck result
$\Omega_{DM} = 0.1199\pm 0.0027$ \cite{Planck}
(in combination with the WMAP data \cite{WMAP}) with a $10\%$ theoretical uncertainty).
\item[(6)] The dark matter in our scenario can be the mixed bino-higgsino. In this case, the direct
detection experiments can possibly set stringent constraints on dark matter. We survey the
spin-independent (SI) direct detection bounds from LUX \cite{LUX}, Xeon1T \cite{XENON1T} and
the future LUX-ZEPLIN 7.2 Ton \cite{LZ} experiment.
\eit

%%fig1 %%%%%%%%%%%%%%%%%%%%%%%
\begin{figure}[ht]
\centering
\includegraphics[width=2.9in]{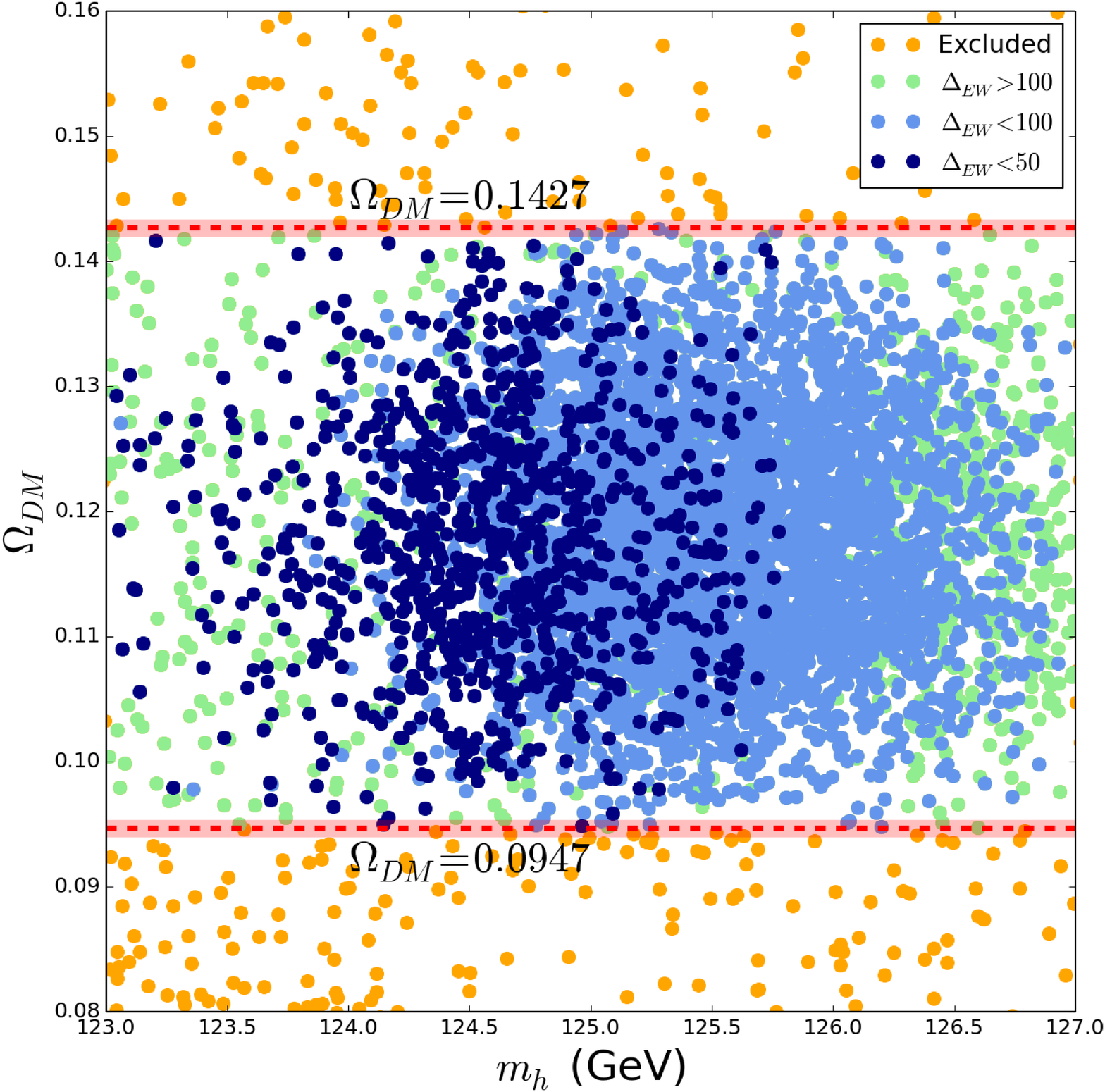}
\includegraphics[width=2.9in]{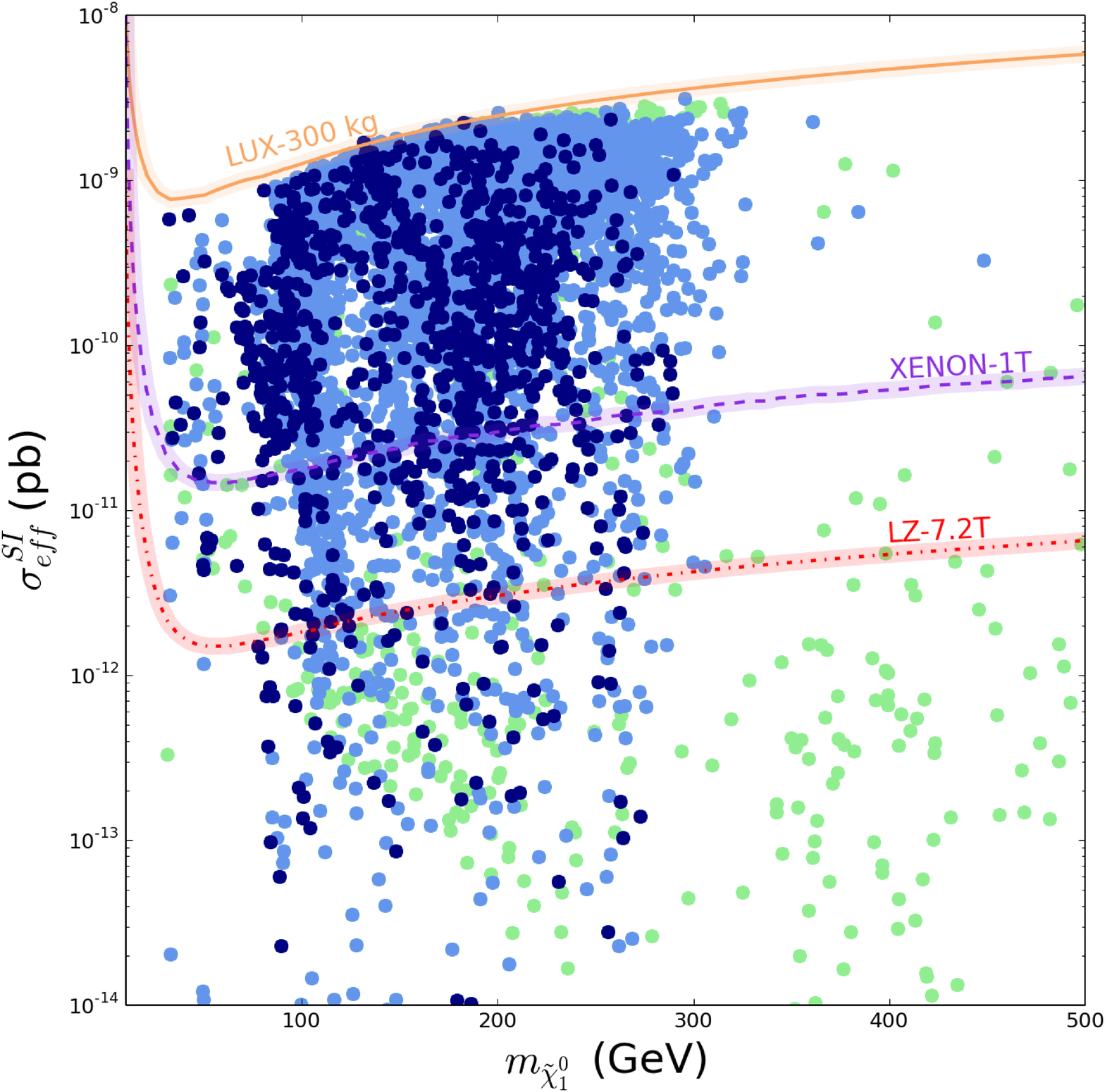}
\vspace{-0.5cm}
\caption{ The scatter plots of the parameter space in our scenario, showing
the dark matter relic density versus the Higgs mass in the  left panel
and the spin-independent DM-nucleon scattering cross section versus the LSP mass
in the right panel.
All the points can survive the collider and dark matter constraints (1-6).
The EW fine tuning ($\Delta_{EW}$) for the sample points are also shown.}
\label{results}
\end{figure}
%%%%%%%%%%%%%%%%%%%%%%%%%

The numerical results with the corresponding EW fine-tuning are shown in Fig 1. It should be noted that\cite{EWFT:over} conventional measures, include BG measure\cite{BGmeasure}, tend to overestimate EWFT in supersymmetric models, often by several orders of magnitude.  Accord to the Fine-tuning Rule proposed in \cite{EWFT:rule}, both Higgs mass and the traditional $\Delta_{BG}$ fine-tuning measures reduce to the model-independent EW fine-tuning measure $\Delta_{EW}$.

From the figure, we have the following observations:
\bit

\item  Both the 125 GeV Higgs mass and the correct DM relic density can be obtained in our scenario.
We can see that there is a large parameter space which can give the correct relic abundance of DM.
This is the consequence of the mixed bino-higgsino DM nature in our scenario. The deflection of
AMSB trajectory is crucial for a light bino to be the lightest gaugino (with $M_1\lesssim \mu$)
that can be compatible with the LHC constraints on gluino mass $m_{\tl{g}} \gtrsim 1.3 {\rm TeV}$.
Without the deflection, the lightest gaugino would be heavy and at the same time wino-like.
Such a relation would predict either higgsino or mixed wino-higgsino DM, both of which
would lead to under-abundance of DM.

\item  Our scenario can also give the observed 125 GeV Higgs mass. This is the consequence of
a relatively large $A_t$ term. Besides, the EW fine-tuning needed for the 125 GeV Higgs mass can be as low as $\Delta_{EW}\lesssim 50$.
 Larger higgs mass will slight increase the EW fine tuning involved.

\item  We also survey the spin-independent (SI) direct detection bounds from DM-nucleon scattering
experiments. It is well known that the SI interaction of the neutralino DM with quarks inside
the nucleus occurs via the $s$-channel squark exchange and $t$-channel Higgs exchange processes.
As squarks are bounded by the LHC data to be considerably heavy, the Higgs exchange diagrams
would dominantly contribute to the spin-independent $\chi-p$ scattering cross section.
The Higgs-$\chi$-$\chi$ coupling is driven by bino-higgsino and wino-higgsino mixing.
Unlike the case for a pure gaugino or a pure higgsino DM  in which the associated SI cross-section
would become quite small, the SI cross section could be large when DM is the mixed bino-higgsino.
However, the DM can evade the SI direct detection experiments if the mixing is small.
In our numerical study, we find that the most interesting points with low EW fine tuning(namely the points that can account for the 125 GeV
Higgs mass with $\Delta_{EW}<100$) have typically a cross section below $10^{-9} pb$. The majority of such points will be covered by XENON-1T. 
On the other hand, there are still small regions with low EW fine tuning that can survive the XENON-1T and LUX-ZEPLIN 7.2 Ton sensitivity. Such points may indicates that the corresponding mixing of bino-higgsino is not large.

\eit

\section{\label{sec-3}Conclusions}
In this work a radiative natural SUSY spectrum were proposed in
deflected anomaly mediation scenario with general messenger-matter interactions. Due to the contributions
from new interactions, positive slepton masses as well as a large $|A_t|$ term can naturally be obtained
with either sign of deflection parameter and few messenger species (thus avoid the possible Landau
pole problem). In this scenario, in contrast to the ordinary radiative natural SUSY scenario with
under-abundance of DM, the DM can be the mixed bino-higgsino and give the right relic density.
The 125 GeV Higgs mass can also be easily obtained in our scenario. The majority of low EW fine tuning points
can be covered by the XENON-1T direct detection experiments.

\section*{Acknowledgement}
We are very grateful to the referee for helpful discussions.
 This work was supported by the National Natural Science Foundation of China (NNSFC)
under grant No. 11105124, 11105125, 11135003 and 11275245,
by the Open Project Program of State Key Laboratory of Theoretical Physics,
Institute of Theoretical Physics, Chinese Academy of Sciences (No.Y5KF121CJ1),
by the Innovation Talent project of Henan Province under grant
number 15HASTIT017, by the Young-Talent Foundation of Zhengzhou University,
and by the CAS Center for Excellence in Particle Physics (CCEPP).

\appendix
\section{Scalar soft SUSY breaking mass terms}
 The expression for the scalar soft parameters are derived from the general forms
in \cite{fei:PLB} and given by
\beqa
\f{m^2}{F_\phi^2}&=&\f{d}{2}\delta^m+\f{d^2}{4}(\delta^G+\delta^3)+\f{1}{4}\delta^A,
\eeqa
with each type of contributions given below.
The relevant expressions are
\bit
\item Cross term (anomaly-gauge mediation) contributions:

The anomaly-gauge mixed mediation part given by
\beqa
\delta^m&=&\f{\pa^2}{\pa\mu\pa\ln |X|}\ln Z^D_{ab}~,\nn\\
&=&(\f{\Delta G^D_a}{2}\f{\pa }{\pa Z_a^D}+\Delta\beta_{g_r}\f{\pa}{\pa g_r})G^--G_a^D\f{\Delta G_a}{2}~.
\eeqa

Cross the messenger threshold, the change of the beta function for $g_i$ is given by
\beqa
\Delta \beta_{g_i}=\f{1}{16\pi^2}N_F g_i^3
\eeqa
and the discontinuity of $G^a$ is
\beqa
\f{\Delta G_{L}}{2}&=&-\f{1}{8\pi^2} \la_L^2~,\nn\\
\f{\Delta G_{E_L^c}}{2}&=&-\f{1}{8\pi^2} 2\la_L^2~,\nn\\
\f{\Delta G_{Q_L}}{2}&=&-\f{1}{8\pi^2} (\la_D^2+\la_U^2)~,\nn\\
\f{\Delta G_{U_L^c}}{2}&=&-\f{1}{8\pi^2} (2\la_U^2)~,\nn\\
\f{\Delta G_{D_L^c}}{2}&=&-\f{1}{8\pi^2} (2\la_D^2)~.
\eeqa
After some manipulations, we can obtain
\beqa
\delta^m_Q&=&\f{1}{8\pi^2}\[y_t^2\f{\Delta G_{y_t}}{2}+y_b^2\f{\Delta G{y_b}}{2}\]+\Delta\beta_{g_r}\f{\pa}{\pa g_r}G_Q^D~,\nn\\
&\approx&-\f{1}{8\pi^2}\[2y_t^2\f{1}{16\pi^2}\(3\la_U^2+\la_D^2\) -2\f{8}{3}\f{1}{16\pi^2}N_F g_3^4\],\nn\\
\delta^m_U&=&-\f{1}{8\pi^2}\[4y_t^2\f{1}{16\pi^2}\(3\la_U^2+\la_D^2\)-2\f{8}{3}\f{1}{16\pi^2}N_F g_3^4\],\nn\\
\delta^m_{D}&=&-\f{1}{8\pi^2}\[-2\f{8}{3}\f{1}{16\pi^2}N_F g_3^4\],\nn\\
\delta^m_{L}&=&\delta^m_E=\delta^m_{H_D}=0~,\nn\\
\delta^m_{H_U}&=&-\f{1}{8\pi^2}\[6y_t^2\f{1}{16\pi^2}\(3\la_U^2+\la_D^2\)\],
\eeqa
Expressions for the first two generation squarks can be obtained by simply removing the $y_t^2$ terms.
\item  Gauge mediation-type contributions:

The gauge mediation part given by
\beqa
\delta^G+\delta^3=-\f{\pa^2}{\pa\ln|X|^2}\ln Z=-\f{\pa^2}{\pa \ln|X|^2} Z+|\f{\pa  Z}{\pa \ln |X|}|^2~.
\eeqa
The sums of the discontinuity are
\beqa
\sum\Delta\(\f{\pa G^{Q}}{\pa Z_a}\)G_a&=&\f{1}{8\pi^2}\[\la_U^2(G_{\la_U}-G_Q)+\la_D^2(G_{\la_D}-G_Q)\]~,\nn\\
\sum\Delta\(\f{\pa G^{U}}{\pa Z_a}\)G_a&=&\f{1}{8\pi^2}\[2\la_U^2(G_{\la_U}-G_U)\]~,\nn\\
\sum\Delta\(\f{\pa G^{D}}{\pa Z_a}\)G_a&=&\f{1}{8\pi^2}\[2\la_D^2(G_{\la_D}-G_D)\]~,\nn\\
\sum\Delta\(\f{\pa G^{L}}{\pa Z_a}\)G_a&=&\f{1}{8\pi^2}\[\la_L^2(G_{\la_L}-G_L)\]~,\nn\\
\sum\Delta\(\f{\pa G^{E}}{\pa Z_a}\)G_a&=&\f{1}{8\pi^2}\[\la_U^2(G_{\la_L}-G_E)\]~,
\eeqa
with $G^U_{\la_U}=G^U_{Q}+G^U_{U}+G^U_{X_u}$ and $G^U_{\la_D}=G^U_{Q}+G^U_{D}+G^U_{X_d}$
the anomalous dimension for $\la_U$ and $\la_D$ above the threshold.
So we can obtain
\beqa
\delta^G_Q&=&\f{1}{8\pi^2}\[y_t^2\f{\Delta G_{y_t}}{2}-\la_U^2G^{TU}_{\la_U}-\la_D^2 G^{TU}_{\la_D}\]~,\nn\\
\delta^G_{U}~&=&\f{1}{8\pi^2}\[2y_t^2\f{\Delta G_{y_t}}{2}-2\la_U^2 G^{TU}_{\la_U}\]~,\nn\\
\delta^G_{D}~&=&\f{1}{8\pi^2}\[-2\la_D^2 G^{TU}_{\la_D}\]~,\nn\\
\delta^G_{L}~&=& \f{1}{8\pi^2}\[-\la_L^2 G^{TU}_{\la_L}\],\nn\\
\delta^G_E~&=&\f{1}{8\pi^2}\[-2\la_L^2 G^{TU}_{\la_L}\]~,\nn\\
\delta^G_{H_D}&=&0~,\nn\\
\delta^G_{H_U}&=&\f{1}{8\pi^2}\[3y_t^2\f{\Delta G_{y_t}}{2}-3\la_U^2 G^{TU}_{\la_U}\]~,
\eeqa
The index TU denotes the value upon the messenger threshold. We list their expressions:
\beqa
\f{\Delta G_{y_t}}{2}&=&-\f{1}{8\pi^2}\(3\la_U^2+\la_D^2\)~,\nn\\
\f{\Delta G_{y_b}}{2}&=&-\f{1}{8\pi^2}\(\la_U^2+3\la_D^2\)~,\nn\\
G^{TU}_{\la_U}&=&-\f{1}{8\pi^2}\(6\la_U^2+\la_D^2+3y_t^2-\f{16}{3}g_3^2\)~,\nn\\
G^{TU}_{\la_D}&=&-\f{1}{8\pi^2}\(6\la_D^2+\la_U^2+\la_L^2+y_t^2-\f{16}{3}g_3^2\)~,\nn\\
G^{TU}_{\la_L}&=&-\f{1}{8\pi^2}\(4\la_L^2+3\la_D^2\)~.
\eeqa

 There are other terms from ordinary GMSB part with
\beqa
\delta^3= \Delta \beta_{g_r}\(\f{\pa}{\pa g_r}G^{TD}\)=\f{1}{8\pi^2} 2c_r 2g_r\f{N_F}{16\pi^2}g_r^3.
\eeqa
Note that the change of the beta function is $\Delta \beta_g=N_F$.
  \beqa
 \delta^3_Q&=& \delta^3_{U}=\delta^3_{D}=\f{N_F}{(8\pi^2)^2}\[\f{8}{3}g_3^4\]~,\nn\\
\delta^3_{L}~&=&\delta^3_E=\delta^3_{H_D}=\delta^3_{H_U}\approx 0.
  \eeqa
In the previous expressions, we keep the terms involving only $g_3$.
\item
Pure anomaly contributions:
\beqa
\delta^A=-\f{\pa^2}{\pa\ln|X|^2}\ln Z=-\f{\pa^2}{\pa \ln|X|^2} Z+|\f{\pa  Z}{\pa \ln |X|}|^2~.
\eeqa
So we obtain
\beqa
\delta_Q^A&=&-\f{1}{8\pi^2}\[y_t^2G_{y_t}+y_b^2G_{y_b}\]-\f{1}{4\pi^2}\[\f{1}{30}b_1\al_1^2+\f{3}{2}b_2\al_2^2+\f{8}{3}b_3\al_3^2\]~,\nn\\
&\approx&\f{1}{(8\pi^2)^2}\[y_t^2\(6y_t^2-\f{16}{3}g_3^2\)\]-\f{1}{4\pi^2}\f{8}{3}b_3\al_3^2~,\nn\\
\delta_U^A&=&-\f{1}{8\pi^2}\[2y_t^2G_{y_t}\]-\f{1}{4\pi^2}\[\f{8}{15}b_1\al_1^2+\f{8}{3}b_3\al_3^2\]~,\nn\\
&\approx&\f{1}{(8\pi^2)^2}\[2y_t^2\(6y_t^2-\f{16}{3}g_3^2\)\]-\f{1}{4\pi^2}\f{8}{3}b_3\al_3^2~,\nn\\
\delta_D^A&\approx&-\f{1}{4\pi^2}\f{8}{3}b_3\al_3^2~,\nn\\\nn\\
\delta_{H_u}^A~&=&\f{1}{(8\pi^2)^2} 3y_t^2\(6y_t^2-\f{16}{3}g_3^2\)~,\nn\\
\delta_L^A~&=&\delta_E^A=\delta_{H_d}^A \approx0~.
\eeqa
\eit

\end{document}